\documentclass{ws-ijmpd}

\begin{document}

\markboth{Maurizio Spurio}
{ANTARES neutrino telescope}

%
\catchline{}{}{}{}{}
%
\title{ANTARES neutrino telescope: status, first results and sensitivity for the diffuse neutrino flux}

\author{Maurizio Spurio, on behalf of the ANTARES collaboration.}
\address{Dipartimento di Fisica dell'Universita' and Sezione INFN di Bologna,\\ 
        viale C. Berti-Pichat,6/2, 40127 - Bologna, Italy\\
        maurizio.spurio@unibo.it}
       
\maketitle

\begin{abstract}
ANTARES is a neutrino telescope under the Mediterranean Sea, in a site 40 km off the French coast at a depth of 2475 m. It is an array of 12 lines equipped with 884 photomultipliers. The detection mechanism relies on the observation of the Cherenkov light emitted by charged leptons produced by neutrinos interacting in the water and ground surrounding the detector.
First studies of the detector performances and preliminary results on reconstruction of atmospheric muons and neutrinos are presented, with the expected sensitivity for a diffuse flux of high energy neutrinos.
\end{abstract}


\section{The ANTARES detector}	
The ANTARES collaboration has completed (May 2008) the construction of the largest neutrino telescope in the Northern hemisphere. It has been installed about 40 km off the coast of Toulon, France, at a maximum depth of 2475 m in the Mediterranean Sea. 
A neutrino telescope in the Northern hemisphere can look at the centre of our Galaxy, and is complementary to the IceCube Antarctic telescope\cite{amanda}. 
The detection principle relies on the detection of the Cherenkov photons emitted by charged leptons produced in neutrino interactions, using a 3D array of photomultipliers (PMTs), in the medium surrounding the detector. 

\begin{figure}[h]
  \begin{center}
\centerline{\psfig{file=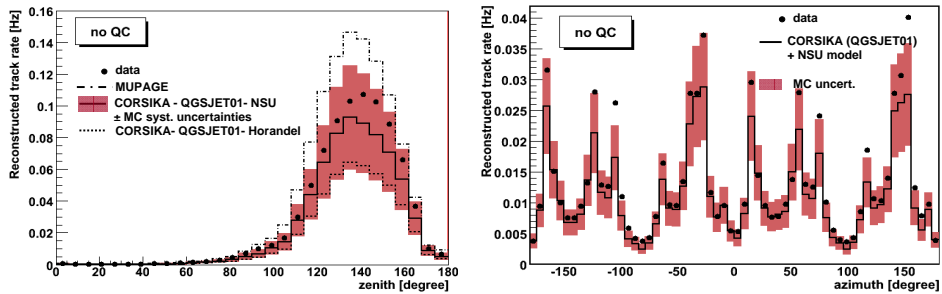,width=12.7cm}}
\caption{\label{fig1} {(a) Zenith and (b) azimuth distributions of reconstructed tracks. Black points represent data. Lines refer to MC expectations, evaluated with two different simulation.  The shadowed  band represents the systematic error due to environmental and geometrical parameters.}}
\end{center}
\end{figure}

The first line was deployed and connected on February 2006. A second line was put into operation in October of the same year. 3 more lines were connected in January 2007 and 5 more, plus the Instrumentated Line, in December 2007. Finally, the apparatus reached its complete configuration when the last two lines were deployed and connected in May 2008.

The detector consists of 12 lines each with a total height of 450 m, set at a distance of 60-70 m from each other, plus an instrumented line to monitor the environment. They are weighted to the seabed and held nearly vertical by buoys at the top. The lines are flexible and can move in the sea current, with displacements of few meters at the top for a typical sea current of 5 cm/s. Each line carries a total of 75 10"-Hamamatsu PMTs housed in glass spheres, the optical modules (OMs) \cite{ant02}, arranged in 25 storeys separated by 14.5 m. A titanium cylinder placed on each storey encloses the electronics for readout and control, together with compasses/tiltmeters used for geometrical positioning. 
Some of the storeys house LED beacons, which are used for timing calibrations \cite{ant07b}. The positions of the OMs are measured with a system of acoustic transponders and receivers at known positions on the line and on the seabed. The positioning system gives a real time measurement, every two minutes, of the OMs position with a precision of $\sim$ 10 cm. 
Each PMT is read-out by 2 ASICs cards (ARS), to reduce dead time, recording time and pulse-height information when a threshold corresponding to 1/3 of a photoelectron is reached. Precise timing is provided by a 20 MHz high accuracy on-shore clock synchronised with the GPS, distributed via the electro-optical cable to each electronics module. 
All digitized information are sent to shore and treated in a computer farm at the shore station where a fast processing of events, satisfying predetermined trigger requirements, is performed \cite{ant07c}.

\section{Expected performances and first results of ANTARES.}
The main ANTARES goal is the detection of neutrinos of cosmic origins, whether coming from discrete sources or due to a diffuse flux. Muon-induced by $\nu$-interactions produce long tracks in the detector which can be reconstructed. An angular resolution better than 0.3$^o$ can be reached above 1 TeV,  the main limitation being the light scattering in water and the electronics. 

Thanks to its modular geometry ANTARES was able to take data in a non complete configuration. In the period February - November 2007, ANTARES run in a 5-line configuration. More than $10^7$ triggers were collected, most of which due to atmospheric muons. They represent the most abundant signal even in a neutrino detector which is optimized for upward-going particle detection. Fig. \ref{fig1} shows the zenith and azimuth distributions of reconstructed muon tracks. Black points represent experimental data. The solid \cite{nsu} and the dotted \cite{horandel} lines refer to Monte Carlo (MC) expectations obtained using a CORSIKA simulation with QGSJET01 for hadronic interaction description, and two CR composition models. The dashed-dotted line refers to a fast simulation using MUPAGE \cite{mupage}, based on parametric formulas of the underwater muon flux \cite{app-para}.
The shadowed band gives an estimate of the systematic errors. This is due to the uncertainties on the environmental parameters, like absorption and scattering lengths of water in ANTARES site, and on the geometrical characteristics of the detector. In particular, given the fact that the OMs are pointing downwards, at an angle of 135$^o$ w.r.t. the vertical, the knowledge of the OMs angular acceptance at these large angles is critical for an accurate determination of the muon flux. 

\begin{figure}[pb]
\centerline{\psfig{file=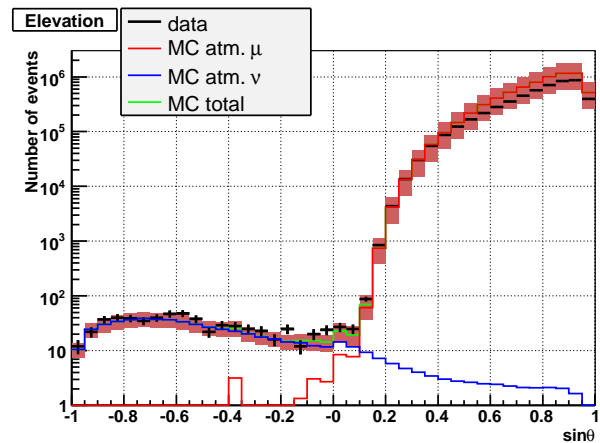,width=8.7cm}}
\caption{Data (black crosses): 173 days of active time with 10 to 12 lines. Blue line: MC simulation of atmospheric neutrinos. Red line: atmospheric muons (Corsika+Horandel). 582 upward going events are found, to be compared to 494 expected from atmospheric neutrinos plus 13 from wrong reconstructed atmospheric muons.}
\label{fig2} 
\end{figure}

A different analysis is necessary when selecting neutrinos. A set of more severe quality cuts must be applied in order to remove  downward-going tracks wrongly reconstructed as upward. Data presented in Fig. \ref{fig2} were collected during the 10-12 line configuration period, from December 2007 to December 2008. The shadowed band represents the sum of theoretical  and systematic uncertainties.

\begin{figure}
\begin{center}
\centerline{\psfig{file=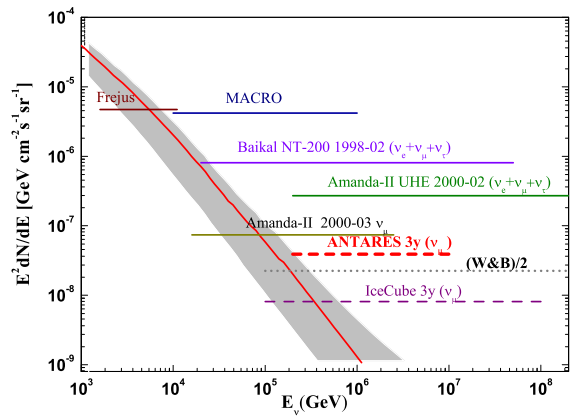,width=8.7cm}}
\vskip -0.7cm\caption{\label{fig3} {Sensitivities and upper limits for a $E^{-2}$ diffuse high energy neutrino flux. Experimental upper limits are indicated as solid lines, the ANTARES and IceCube 90\% C.L. sensitivities with dashed lines. The Frejus$^{13}$, MACRO$^{14}$, Amanda-II 2000-03$^{16}$ limits refers to on $\nu_\mu$. The Baikal$^{15}$ and Amanda-II  UHE 2000-02$^{17}$ refers to neutrinos of all-flavors.  For reference, the Waxman and Bahcall limit for transparent sources it is also shown.
 It was divided by two, to take into account neutrino oscillations.}}
\end{center}
\end{figure}

The results for the search for point-like sources is presented in \cite{ciro}. The search of unresolved (neither in time nor in location) neutrino sources (diffuse flux) is based on the observation of an excess of high energy events above the
irreducible background of atmospheric neutrinos. MC studies of the  sensitivity, show that ANTARES can constrain theoretical models on extragalactic neutrino production and reach the level of few times the WB\cite{WB} upper bound in 3 years of operation (see Fig. \ref{fig3}).

\section{Conclusions}
The ANTARES neutrino telescope is running in its final configuration of 12 detection lines, after a long period of development and construction. 
The first data demonstrate that the detector performances are within the expectations\cite{line1} , in particular for the possibility of a good measurement of the neutrino direction. In addition, the results obtained until now allowed to check the reconstruction programs and MC simulations. 
The angular distributions of atmospheric downgoing muons and of atmospheric neutrinos measured with 5 lines are in agreement within 30 - 40\% in absolute normalization with MC expectations, the main source of uncertainty being the primary cosmic ray flux and interaction model and the angular acceptance of the OMs.\\
ANTARES results show as well that used technology is reliable and mature to carry out the project of a km$^3$ neutrino telescope in the Mediterranean Sea \cite{km3net}.
\vskip -1.0cm

\vskip -1.7cm


\end{document}